\documentclass[sigconf]{acmart} 

\usepackage{soul}
\usepackage{graphicx}
\usepackage{booktabs}
\usepackage{float}
\usepackage{overpic}
\usepackage{float}  
\usepackage{subfloat}
\usepackage{stfloats} 
\usepackage[skip=0.5\baselineskip]{caption}
\usepackage{bm}
\usepackage{amsmath}
\usepackage{booktabs}
\usepackage{multirow}
\usepackage{multicol}
\usepackage{enumitem}
\usepackage{subcaption}
\usepackage{threeparttable}
\usepackage{xspace}
\usepackage{color}
\usepackage[normalem]{ulem}
\usepackage{algorithmicx,algorithm}
\usepackage{algpseudocode}
\usepackage{algorithm,algpseudocode}
\usepackage{color, xspace}

\AtBeginDocument{%
  \providecommand\BibTeX{{%
    \normalfont B\kern-0.5em{\scshape i\kern-0.25em b}\kern-0.8em\TeX}}}

\acmISBN{978-1-4503-XXXX-X/18/06}

\begin{document}

\title{Joint Modeling in Recommendations: A Survey}
\author{Xiangyu Zhao$^1$, Yichao Wang$^2$, Bo Chen$^2$, Jingtong Gao$^1$, Yuhao Wang$^1$}
\author{Xiaopeng Li$^1$, Pengyue Jia$^1$, Qidong Liu$^{1,3}$, Huifeng Guo$^2$, Ruiming Tang$^2$}
\affiliation{
	\institution{$^1$City University of Hong Kong, $^2$Huawei Noah’s Ark Lab, $^3$Xi'an Jiaotong University}
	\country{}
}
\email{xianzhao@cityu.edu.hk, {jt.g,yhwang25-c,xiaopli2-c,jia.pengyue}@my.cityu.edu.hk}
\email{liuqidong@stu.xjtu.edu.cn, {wangyichao5,chenbo116,huifeng.guo,tangruiming}@huawei.com}

\renewcommand{\shortauthors}{Xiangyu Zhao, et al.}

\begin{abstract}
In today's digital landscape, Deep Recommender Systems (DRS) play a crucial role in navigating and customizing online content for individual preferences. However, conventional methods, which mainly depend on single recommendation task, scenario, 
data modality and user behavior, are increasingly seen as insufficient due to their inability to accurately reflect users' complex and changing preferences. This gap underscores the need for joint modeling approaches, which are central to overcoming these limitations by integrating diverse tasks, scenarios, modalities, and behaviors in the recommendation process, thus promising significant enhancements in recommendation precision, efficiency, and customization.
In this paper, we comprehensively survey the joint modeling methods in recommendations. We begin by defining the scope of joint modeling through four distinct dimensions: multi-task, multi-scenario, multi-modal, and multi-behavior modeling. Subsequently, we examine these methods in depth, identifying and summarizing their underlying paradigms based on the latest advancements and potential research trajectories. Ultimately, we highlight several promising avenues for future exploration in joint modeling for recommendations and provide a concise conclusion to our findings.
\end{abstract}




\maketitle

\section{Introduction}
In the modern digital ecosystem, the exponential growth of online content has precipitated a significant challenge for users: Information overload~\cite{aljukhadar2010information}. This predicament has catalyzed the development and widespread adoption of Deep Recommender Systems (DRSs)~\cite{zhang2019deep,batmaz2019review}, which have become foundational to various online platforms, providing tailored content suggestions to streamline user experiences. The traditional DRS models follow the simple input-output deep learning framework, starting from the processing of input data, going through the embedding table of dimensionality reduction, participating in feature interaction, and finally generating the recommendation results. This sequence has proven to be effective in mitigating the challenges posed by the vast seas of digital information, thereby personalizing the digital landscape for users according to their preferences and historical interactions~\cite{xue2017deep,cheng2016wide}.

Despite the notable successes of traditional DRSs, they are not devoid of shortcomings. Primarily, these systems are characterized by a single-task orientation~\cite{tang2020progressive} that narrows the horizon of information extraction. Furthermore, the approach focusing on single-scenario~\cite{zhu2021cross,wang2022causalint} overlooks the synergy between different scenarios and also leads to excessive resource consumption as it maintains a specific model for each scenario. Another significant limitation is their reliance on ID-based (single modality) features, which restricts the scope of information that can be extracted from data samples~\cite{wei2023multi}. Last but not least, the lack of a comprehensive strategy for understanding complex relationships through user behaviors hinders the system's ability to provide nuanced and context-sensitive recommendations~\cite{yang2022multi}.

In response to these limitations, the concept of joint modeling has been introduced and is gaining momentum as a powerful enhancement of the traditional DRS frameworks. This innovative approach, which encompasses multi-task~\cite{wang2019multi,tang2020progressive}, multi-scenario~\cite{sheng2021one,yang2022adasparse}, multi-modal~\cite{wei2019mmgcn,chen2017attentive}, and multi-behavior~\cite{jin2020multi,xia2021knowledge} 
strategies, aims to transcend the conventional boundaries of recommender systems. Joint modeling endeavors to leverage a broader spectrum of data modalities and user behaviors, implement more sophisticated modeling techniques, and streamline the deployment process across diverse scenarios and tasks. This approach is designed to facilitate the delivery of more accurate, personalized, and contextually appropriate recommendations, thereby improving user experience.

Specifically, joint modeling in the recommender systems mainly includes the following dimensions. 
\begin{itemize}[leftmargin=*] 
    \item \textbf{Multi-task Modeling}: The multi-task recommendation paradigm within joint modeling seeks to amalgamate and exploit synergies between different recommendation tasks, thereby enhancing overall efficiency and output quality. Generally, the advantage of multi-task lies in good generalization or robustness.
    \item \textbf{Multi-scenario Modeling}: The multi-scenario recommendation framework aims to harness the rich diversity of data across various user interactions, unifying disparate scenario-specific models to improve system adaptability and robustness. Generally, multi-scenario resolved the sparsity and improved the efficiency.
    \item \textbf{Multi-modal Modeling}: Multi-modal recommender systems strive to incorporate an array of data types--including textual, visual, and auditory information--to construct a more comprehensive and nuanced understanding of user preferences. 
    \item \textbf{Multi-behavior Modeling}: These recommendation strategies offer a more granular analysis of user behaviors. This enables the capturing of intricate user interaction patterns, thereby paving the way for a new generation of recommender systems capable of delivering unparalleled personalization and relevance.
\end{itemize}

\begin{figure*}[t]
    \centering
    \includegraphics[width=1.0\linewidth]{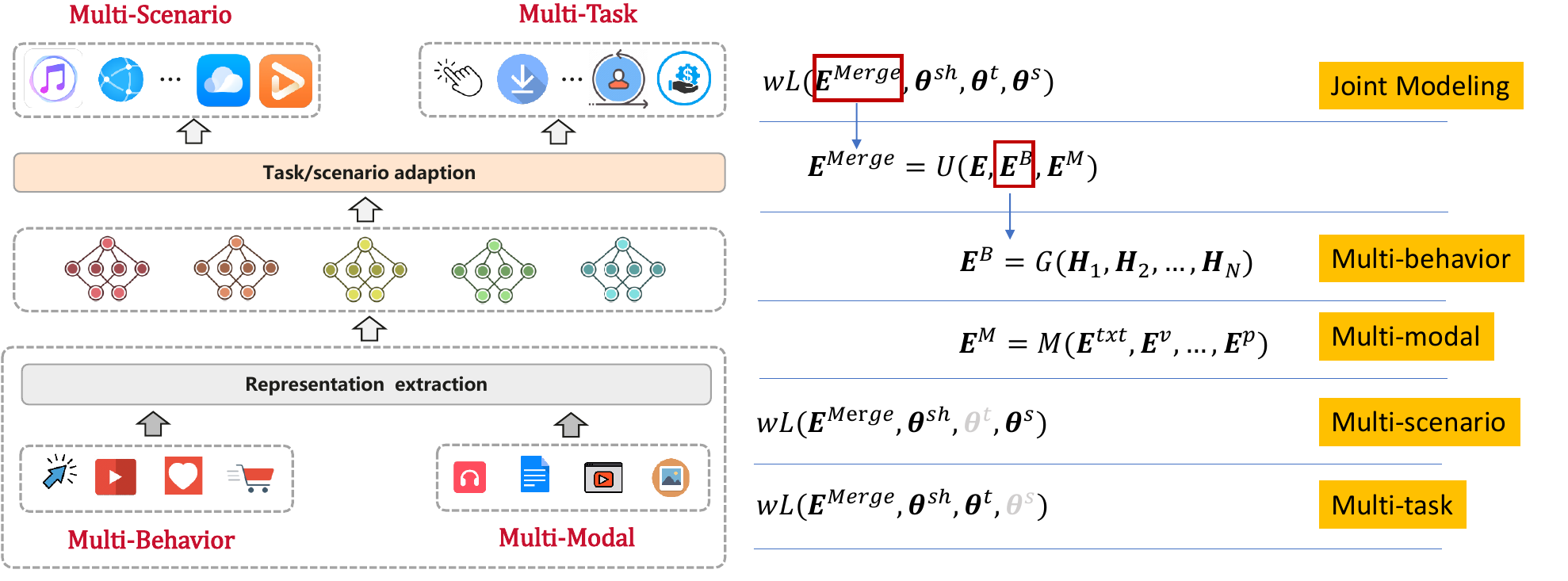}
    \caption{Joint modeling in recommendations.}
    \label{fig:problem}
    \vspace{-5mm}
\end{figure*}

The overall goal of this study is to thoroughly explore and explain the growing field of joint modeling in recommender systems. This work aims to understand the complex details, review the latest methods, and highlight the ongoing issues and new opportunities for further research. Initially, the paper outlines the basic problem formulation of traditional recommender systems and joint modeling approaches. Then, it goes on to examine each dimension of joint modeling in detail, clarifying its importance, taxonomy, and current research priorities. Later, the study looks at possible directions for the future of joint modeling. Finally, this paper summarizes the role of joint modeling in improving recommender systems and stresses the need for further innovation in this area, presenting it as crucial for the advancement of these systems. By doing so, this study seeks to chart a course for the evolution of recommender systems, guiding the incorporation of joint modeling techniques to foster the development of more sophisticated and effective platforms.

\section{Problem Formulation}
Considering a recommender system modeled by a function $f(\cdot)$, depending on a single recommendation task and scenario, single data
modality, or single user behavior, aiming to optimize the model parameters $\boldsymbol{\theta}$ through a specified loss function $L(\boldsymbol{\theta})$. Diverging from this, the joint modeling approach facilitates refined analyses across multiple dimensions, including multi-task, multi-scenario, multi-modal, and multi-behavior modeling, as depicted in Figure~\ref{fig:problem}. 
Specifically, in the joint modeling frameworks, \underline{\textbf{multi-behavior}} and \underline{\textbf{multi-modal}} modeling could be applied to encode user behavior sequences and multi-modal data into dense representations: 
\begin{equation}
\begin{aligned}
\boldsymbol{E}^{B}=&G\left(\boldsymbol{H}_1, \boldsymbol{H}_2, \ldots, \boldsymbol{H}_N\right), \\
\boldsymbol{E}^M=&M\left(\boldsymbol{E}^{t x t}
\boldsymbol{E}^v, \ldots, \boldsymbol{E}^p\right)
\end{aligned}
\end{equation}
where $E^{B}$ is the behavior representation learned from the user behavior sequence $\left(\boldsymbol{H}_1, \boldsymbol{H}_2, \ldots, \boldsymbol{H}_N\right)$ with $N$ behaviors and $\boldsymbol{E}^M$ is the modality representation learned from different modality input $\left(\boldsymbol{E}^{t x t}, \boldsymbol{E}^v, \ldots, \boldsymbol{E}^p\right)$ with $txt$ for text modality, $v$ for visual modality and $p$ for other possible modality. 
By aggregating the original ID-based representation $\boldsymbol{E}$ and the generated $\boldsymbol{E}^{B}$ and $\boldsymbol{E}^M$, $\boldsymbol{E}^{Merge}$ can aggregate more useful structured information for model learning, and it is thus treated as the final representation of input information for further modeling. 
Additionally, the \underline{\textbf{multi-task}} modeling paradigm could be applied to learn $K$ task relations via 
\begin{equation}
Loss = \sum_{k=1}^K w^k L^k\left(\boldsymbol{E}^{Merge}, \boldsymbol{\theta}^{sh}, \boldsymbol{\theta}^k\right)
\end{equation}
and \underline{\textbf{multi-scenario}} can learn scenario relations between $S$ scenarios via 
\begin{equation}
Loss = \sum_{s=1}^S w^s L^s\left(\boldsymbol{E}^{Merge}, \boldsymbol{\theta^{sh}}, \boldsymbol{\theta}^s\right)
\end{equation}

By dividing parameters for shared structure modeling ($\boldsymbol{\theta}^{sh}$), multi-task modeling ($\boldsymbol{\theta}^k$), and multi-scenario modeling ($\boldsymbol{\theta}^s$), the DRS is capable of handling multiple tasks and scenarios simultaneously and further enhancing its performance. The overall loss function could be a weighted combination: 
\begin{equation}
Loss = \sum_{i=1}^I w^i L^i\left(\boldsymbol{E}^{Merge}, \boldsymbol{\theta}^{sh}, \boldsymbol{\theta}^k, \boldsymbol{\theta}^s\right)
\end{equation}
where $w^i$ is the weight for the $i$-th loss funtion and $I=K+S$.
\section{Multi-Task Recommendation}

In practice, industrial recommender systems should be endowed with the capability to conduct various recommendation tasks simultaneously so as to cater to multiple and diverse demands of users and make profits. Consequently, Multi-task Recommendation (MTR) arises because it offers two main benefits. On the one
hand, it can achieve mutual enhancement among the tasks by exploiting data and knowledge across multiple tasks.
On the other hand, A higher efficiency of computation and storage can be obtained. Specifically, for example, the score prediction task aims to predict the likelihood of a user performing an action, such as click-through rate (CTR) prediction, and the generation task focuses on providing explanations for recommendations.

The objective is to learn the MTL model with task-specific parameters $\{\boldsymbol{\theta}^1,...,\boldsymbol{\theta}^K\}$ and shared parameter $\boldsymbol{\theta}^{sh}$, which outputs the $K$ task-wise predictions by extracting the hidden pattern of user-item feature interactions. The loss function for multi-task training is commonly defined as a weighted sum of losses and can be written as the following optimization problem:
\vspace{-2mm}
\begin{equation} \label{eq:iwl}
    \begin{aligned}
        \underset{\left\{\boldsymbol{\theta}^1,...,\boldsymbol{\theta}^K \right\}}{\arg \min} \mathcal{L}(\boldsymbol{E}^{ {Merge }},\boldsymbol{\theta}^{sh}, \boldsymbol{\theta}^1,\cdots,\boldsymbol{\theta}^K) \\
        = \underset{\left\{\boldsymbol{\theta}^1,...,\boldsymbol{\theta}^K \right\}}{\arg \min} \sum_{k=1}^{K} w^k L^k(\boldsymbol{E}^{{Merge }}, \boldsymbol{\theta}^ {sh}, \boldsymbol{\theta}^k)
    \end{aligned}
\end{equation}
where $L^k(\boldsymbol{\theta}^{sh}, \boldsymbol{\theta}^k)$ is the loss function for the $k$-th task with parameter $\boldsymbol{\theta}^{sh},\boldsymbol{\theta}^k$, and $w^k$ is the loss weight for the $k$-th task.

\subsection{Task Relation}
\begin{figure}
    \centering
    \includegraphics[width=0.7\linewidth]{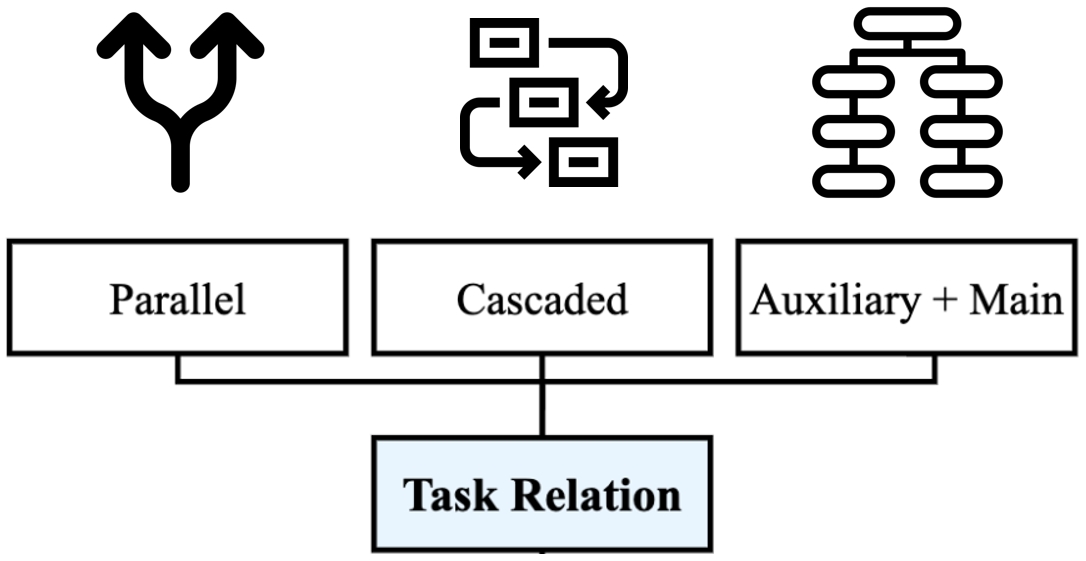}
    \vspace{-1mm}
    \caption{Overview of three task relations: parallel, cascaded, and auxiliary with main in multi-task recommendation.}
    \label{fig:task}
    \vspace{-5mm}
\end{figure}

Specifically, considering the relationship among tasks, multi-task recommendations can be categorized into three groups: parallel, cascaded, and auxiliary with main tasks as shown in Figure \ref{fig:task}. 

\subsubsection{\textbf{Parallel}}

A parallel task relation indicates that various tasks are independently calculated without the sequential dependency of their results.
The objective function of parallel task relation in multi-task recommendation is usually defined as the weighted sum of losses with constant loss weights.
Existing methods under parallel task relation can be grouped by their target and challenge.

As for the target, RnR \cite {hadash2018rank} merges the ranking and rating prediction tasks along with a two-phase decision process for personalized recommendation in video recommendation.
Additionally, MTER \cite{wang2018explainable} and CAML \cite{chen2019co} focus on recommendation and explanation tasks. 
Besides, DINOP \cite{xin2019multi} is proposed specifically for e-commerce online promotions by considering multiple sales prediction tasks.

Meanwhile, several works try to tackle the challenge of feature selection and sharing among parallel tasks, since a static sharing strategy may fail to capture the complex task relevance.
Existing studies mainly adopt attention mechanisms.
DUPN \cite{ni2018perceive} integrates multi-task learning, attention along with RNNs to extract general features that will be shared among the associated tasks.
MRAN \cite{zhao2019multiple} proposes to use an attention mechanism for feature interaction and task-feature alignment.
RevMan \cite{li2021revman} uses an attentive adaptive feature sharing mechanism for different tasks.
MSSM \cite{ding2021mssm} applies a feature-field sparse mask within the input layer and the connection control between a set of more fine cells in sub-networks of multiple layers.
Recently, CFS-MTL \cite{chen2022cfs} proposes to select the stable causal features via pseudo-intervention from a causal view.


\subsubsection{\textbf{Cascaded}} \label{casc}
A cascaded task relationship refers to the sequential dependency between tasks. In other words, the computation of the current task depends on the previous ones, e.g., CTCVR derived by multiplying CTR and CVR. It can be considered as a general MTL problem with the assumption on the prediction scores for each specific task $k$:
\begin{equation}
    \hat{y}_n^k(\boldsymbol{\theta}^{sh}, \boldsymbol{\theta}^k) - \hat{y}_n^{k-1}(\boldsymbol{\theta}^{sh}, \boldsymbol{\theta}^k) = P(\epsilon_k=0, \epsilon_{k-1}=1)
\end{equation}
where $\epsilon_k$ is the indicator variable for task $k$. This assumption implies the difference between $\hat{y}_n^k(\boldsymbol{\theta}^ {sh},\boldsymbol{\theta}^k)$ and $\hat{y}_n^{k-1}$ is the probability of the task $k$ not happening while the task $k-1$ is observed. Besides, the formulation above is equivalent to the sequential dependence MTL (SDMTL) which was raised in \cite{tao2023task}.

The methods under the cascaded task relation setting \cite{ma2018entire,wen2020entire,zhang2020large,wang2020delayed,wen2021hierarchically,xi2021modeling,wu2022multi,wang2022escm2,jin2022multi,tao2023task,zhu2023dcmt} basically aim at CVR prediction task on e-commerce, except AITM \cite{xi2021modeling} and APEM \cite{tao2023task} are proposed for advertising and financial service, respectively. Meanwhile, they mainly target at tackling sample selection bias (SSB) and data sparsity (DS) issues caused by sparse training data of conversion. Besides, their assumed sequential patterns are based on ``impression $\rightarrow$ click $\rightarrow$ conversion'' and its extension following the setting and framework of ESMM \cite{ma2018entire}, which adopts shared embedding and models over the entire space.

\subsubsection{\textbf{Auxiliary with Main Task}} \label{atl}
Auxiliary with main task refers to the circumstance that a task is specified as the main task while others, i.e., associated auxiliary tasks help to improve its performance.
The probability estimation for the main task is calculated based on the probability of auxiliary tasks, which is estimated on the entire space with richer information.

On the one hand, some works simply adopt the original recommendation tasks as auxiliaries \cite{zhang2020large,wang2020delayed,zhao2021distillation,he2022metabalance}. Specifically, Multi-IPW and Multi-DR \cite{zhang2020large} introduce an auxiliary CTR task with the main CVR and imputation task. ESDF \cite{wang2020delayed} treats CTR and CTCVR as auxiliaries of time delay tasks. DMTL \cite{zhao2021distillation} models CTR as an auxiliary of duration task. 

On the other hand, some works design various auxiliary tasks under specific settings \cite{li2020multi,lin2022personalized,yang2021multi,yang2022cross}. Specifically, MTRec \cite{li2020multi} takes link prediction for network dynamic modeling as an auxiliary of the recommendation task. PICO \cite{lin2022personalized} considers task relevance between CTR and CVR as an auxiliary. MTAE \cite{yang2021multi} predicts the winning probability as an auxiliary. Cross-Distill \cite{yang2022cross} proposes a ranking-based task as an auxiliary containing cross-task relation. Specially, CSRec \cite {bai2022contrastive} and PICO \cite{lin2022personalized} adopt contrastive learning as the auxiliary task to extract task relevance better.

Nevertheless, the frameworks above are manually-auxiliary, since the design of auxiliary tasks usually requires specific domain knowledge. Recently, Wang et al. \cite{wang2022can} propose under-parameterized self-auxiliaries to achieve better generalization.

\subsection{Optimization}

From the optimization aspect, the joint optimization of different tasks 
needs to tackle two issues in MTL: (i) the conflict across the performance of multiple tasks, e.g., accuracy, and (ii) the trade-off between objectives in each task. The former is related to negative transfer, while the latter corresponds to a multi-objective trade-off.



\subsubsection{\textbf{Negative Transfer}} \label{nt}

Negative transfer is a situation in which transferring unrelated information among tasks could result in performance degradation and a seesaw phenomenon. When the performance of some tasks is improved but at the cost of others' results, the seesaw phenomenon is observed. Since most of the existing methods seek better recommendation accuracy, they focus on this problem. Specifically, there are mainly two reasons causing negative transfer in MTR from the shared parameters $\boldsymbol{\theta}$. The first is gradient dominating and the second is parameter conflict.

Gradient dominating denotes the magnitude imbalance of gradient $\Vert \nabla_{\boldsymbol{\theta}} L^k(\boldsymbol{\theta}) \Vert$ of different tasks, and some works try to tackle this problem \cite{chen2018gradnorm,yu2020gradient}. In the recommendation community, AdaTask \cite{yang2022adatask} proposes quantifying task dominance of shared parameters and calculating task-specific accumulative gradients for adaptive learning rate methods. 
MetaBalance \cite{he2022metabalance} proposes to flexibly balance the gradient magnitude proximity between auxiliary and target tasks by a relax factor.

Besides, parameter conflict indicates that the shared parameter $\boldsymbol{\theta}$ has opposite directions of gradient $\nabla_{\boldsymbol{\theta}} L^k(\boldsymbol{\theta})$ in different tasks.
PLE \cite{tang2020progressive} discusses the seesaw phenomenon and proposes Customized Gate Control (CGC) that separates shared and task-specific experts to explicitly alleviate parameter conflicts.
CSRec \cite{bai2022contrastive} applies an alternating training procedure and contrastive learning on parameter masks to reduce the conflict probability.

\subsubsection{\textbf{Multi-objective Trade-off}} \label{mt}

The trade-off among objectives under the MTR setting is a new topic. Specifically, the corresponding objectives in each task are usually optimized by a single model regardless of the potential conflict.
Some study the trade-off between group fairness and accuracy across multiple tasks \cite{wang2021understanding} and afterward, the trade-off between minimizing task conflicts and improving multi-task generalization in a higher level \cite{wang2022can}.


\subsection{Training Mechanism} \label{md}
Training mechanism refers to the specific training process and learning strategy of different tasks in the MTR model. Existing works on MTR can be grouped into joint training, reinforcement learning, and auxiliary task learning.


\subsubsection{\textbf{Joint Training}}

Most MTL models adopt joint training among tasks in a parallel manner, and the majority of the above-mentioned MTR models belong to this category. Specially, some works simply jointly learn different tasks, 
such as session-based RS \cite{shalaby2022m2trec,qiu2021incorporating,meng2020incorporating}, route RS \cite{das2022marrs}, knowledge graph enhanced RS \cite{wang2019multi}, explainability \cite{lu2018like,wang2018explainable}, and graph-based RS \cite{wang2020m2grl}. Besides, some works adopt an alternating training procedure, e.g., contrastive pruning \cite{bai2022contrastive}.


\subsubsection{\textbf{Reinforcement Learning}} \label{rl}

Reinforcement Learning (RL) algorithms have recently been applied in DRS, which models the sequential user behaviors as Markov Decision Process (MDP) and utilizes RL to generate recommendations at each decision step \cite{mahmood2007learning}. By setting user-item features as state and continuous score pairs for multiple tasks as actions, the RL-based MTL method is capable of handling the sequential user-item interaction and optimizing long-term user engagement. Zhang et al. \cite{zhang2022multi} formulate MTF as MDP and use batch RL to optimize long-term user satisfaction. Han et al. \cite{han2019optimizing} use an actor-critic model to learn the optimal fusion weight of CTR and the bid rather than adopting greedy ranking strategies to maximize the long-term revenue. Liu et al. \cite{liu2023multi} use dynamic critic networks to adaptively adjust the fusion weight considering the session-wise property.


\subsubsection{\textbf{Auxiliary Task Learning}}

As discussed in Section \ref{atl}, adding auxiliary tasks aims at helping to enhance the performance of primary tasks. Specifically, the auxiliary tasks are usually trained along with the primary tasks in a joint training manner. By contrast, ESDF \cite{wang2020delayed} employs Expectation-Maximization
(EM) algorithm for optimization. Besides, self-auxiliaries are trained with task-specific sub-networks while they are discarded in the inference stage.
\section{Multi-Scenario Recommendation}

\begin{figure*}
    \centering
    \includegraphics[width=0.8\linewidth]{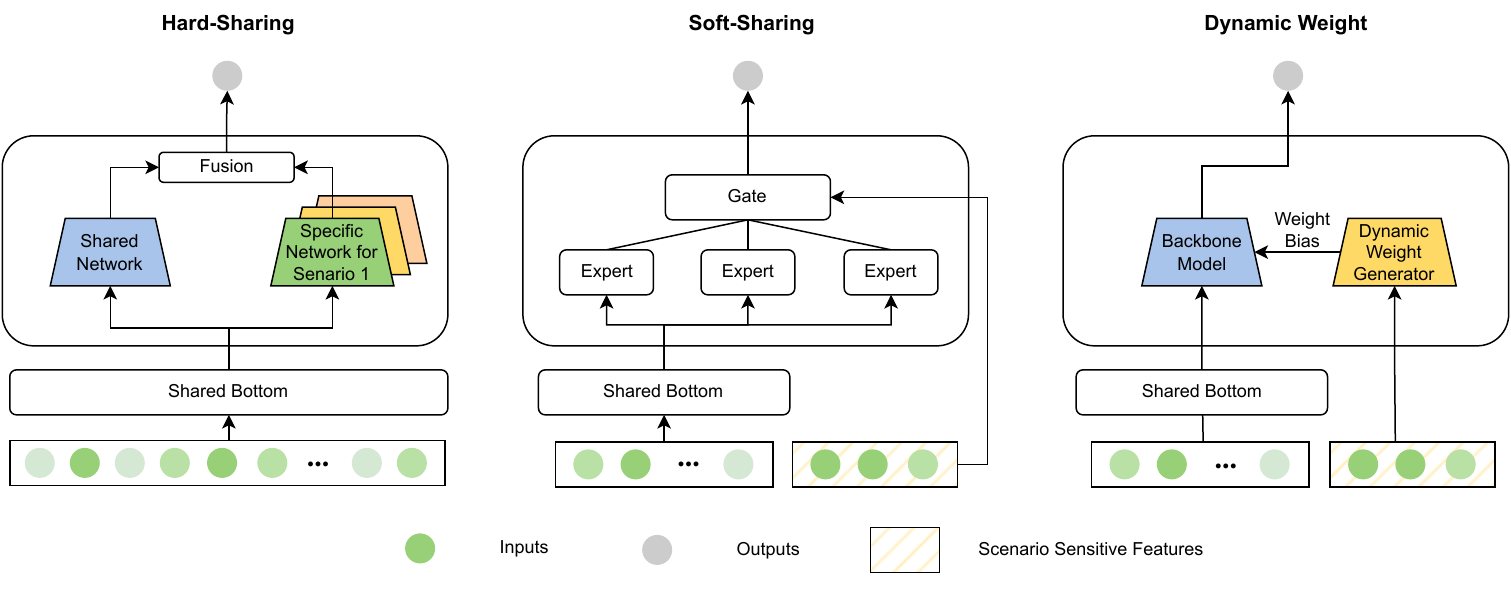}
    \vspace{-3mm}
    \caption{Overview of three categories of multi-scenario modeling methods in recommender systems.}
    \label{fig:overview_msr}
    \vspace{-5mm}
\end{figure*}

Industrial recommender systems are often required to cater to diverse business needs, such as providing personalized recommendations on the homepage and expanding user interests on the detail page. Data from different scenarios share commonalities and diversities. For the commonality, there is an intersection of users and items across different scenarios, resulting in similar data distribution. For the diversities, users behave distinctly when facing different scenarios, which causes diverse data distribution. Traditional approaches typically follow two main strategies. 1) Training separate models for each individual scenario. This method significantly increases the costs of model training and maintenance and also overlooks the commonalities across different scenarios. 2) Simply using all the data from various scenarios to train a unified model. However, this approach fails to consider the differences between scenarios, resulting in subpar performance. 

In light of this context, the concept of Multi-Scenario Recommendation~(MSR) has been introduced. It aims to enhance the recommendation performance for multiple scenarios simultaneously by utilizing a unified model that can effectively capture the commonalities and diversities across different scenarios. As shown in Figure~\ref{fig:overview_msr}, there are mainly three categories of modeling methods in multi-scenario recommendations: hard-sharing modeling, soft-sharing modeling, and dynamic weight paradigm. We will detail these three categories in the following subsections. 

The objective of MSR can be formalized as follows:
\begin{equation} \label{eq:multi-scenario-equation}
    \begin{aligned}
        \arg \min \sum_{s=1}^{S} \boldsymbol{w}^s L^s(\boldsymbol{E}^{Merge}, \boldsymbol{\theta}^{sh}, \boldsymbol{\theta}^s)
    \end{aligned}
\end{equation}
where $\boldsymbol{\theta}^s$ denotes the parameters for each scenario and $\boldsymbol{\theta}^{sh}$ is the shared parameters across different scenarios. $L^s$ is the loss function for $s$-th scenario, $\boldsymbol{w}^s$ is the weights of different loss for $s$-th scenario, and $\boldsymbol{E}^{Merge}$ denotes the merge of all input and generated feature representations. In multi-scenario recommendations, $\boldsymbol{w}^s$ usually is 1, and $L^s$ usually keeps the same across scenarios.

\subsection{Hard-sharing Modeling}

Hard-sharing is a modeling paradigm that was initially proposed in the context of MSR. It primarily focuses on controlling the independence and sharing of network structures or parameters across different scenarios to capture the differences and commonalities between them. STAR~\cite{sheng2021one} is the first paper to propose the hard-sharing method in MSR. It presents a star topology structure, where all data can be transmitted through a shared network. In contrast, specific networks for each scenario only allow data belonging to that particular scenario to pass through. ADI~\cite{jiang2022adaptive} proposes a domain interest adaptation layer to weigh features adaptively and the domain-specific and shared networks to capture the commonalities and diversities across scenarios. AdaptDHM~\cite{li2022adaptdhm} follows the paradigm and introduces a distribution adaptation module to determine the cluster of instances. Then, the relationships between clusters are captured by the multi-distribution network. MMN~\cite{MMN} uses independent weight matrices to model the differences between network designs in different scenarios. Additionally, a unified base parameter is used to model the common characteristics between different scenarios. In addition, EDAA~\cite{EDAA} and Adasparse~\cite{yang2022adasparse} model the commonalities and diversities between scenarios from the perspectives of embedding and pruner, where the embeddings and pruners are isolated across different scenarios.

Hard-sharing in MSR models decouples the commonalities and diversities between scenarios. By controlling the independence and sharing of network structures or parameters across different scenarios, it effectively models the diversities and commonalities between these scenarios. This groundbreaking approach has significantly advanced the field of multi-scenario modeling, allowing for a more nuanced understanding of the dynamics between different contexts. Its innovative contribution has paved the way for further developments in this area.

\subsection{Soft-sharing Modeling}
Unlike hard-sharing, which directly shares model structures or parameters among scenarios, soft-sharing models employ an implicit method, where inter-scenario information is shared through auxiliary models such as cells, layers, or methodologies, including embeddings and loss functions et al. Consequently, we categorize the current methods into two categories: soft-module methods and soft-strategies methods. (1). \textbf{Soft-module methods}, involving leveraging expert networks that construct scenario-specific and scenario-shared experts alongside dynamic routing to integrate information seamlessly. For example, AESM\textsuperscript{2}~\cite{zou2022automatic} introduces an innovative expert selection algorithm to automatically identify scenario-specific and scenario-shared experts for the input. MARIA~\cite{tian2023multi} employs a Mixture of Experts~(MoE) structure with a scenario-shared tower for final predictions, while HiNet~\cite{zhou2023hinet} implements Sub-Experts Integration~(SEI) modules to model scenario-related information, drawing inspiration from MoE. Additionally, SAMD~\cite{huan2023samd} utilizes a meta-network to establish inter-scenario relations, and MAGRec~\cite{ariza2023exploiting} introduces a domain contextualization model that leverages a dense Transformer layer to extract domain representations. 
(2). \textbf{Soft-strategies methods}, focus on deploying various soft strategies to address challenges across scenarios. From the perspective of embedding, EDDA~\cite{ning2023multi} introduces Embedding Disentangling (ED), which segregates inter- and intra-scenario information into distinct embeddings for each item or user. MAMDR~\cite{luo2023mamdr} proposes Domain Negotiation~(DN) to mitigate domain conflict problems using different gradients, and Domain Regularization~(DR) to optimize domain-specific parameters.

\subsection{Dynamic Weight Paradigm}
Dynamic weight generation, a method that dynamically generates weight matrices or parameters for models, has recently gained significant attention for its ability to adapt models to various domains efficiently. For instance, PEPNet~\cite{chang2023pepnet}, developed by Kuaishou, leverages EPNet to map input domain features through a Gate Neural Unit~(GateNU) to produce domain-specific weights. Similarly, M2M~\cite{zhang2022leaving} focuses on generating weight parameters tailored to meta units embedded with scenario-specific knowledge. HAMUR~\cite{li2023hamur} introduces domain-specific weights for adapters plugged into the backbone network, enabling efficient adaptation in various domains. Meanwhile, MI-DPG~\cite{cheng2023mi} generates scenario-conditioned weight matrices via a shallow neural network, which are then decomposed into multiple low-rank matrices to enhance robustness.

\subsection{Multi-scenario Multi-task modeling}
Multi-scenario Multi-task modeling, which integrates perspectives from both Multi-task and Multi-scenario frameworks, aims to achieve optimal performance across various tasks in different scenarios. For instance, M2M~\cite{zhang2022leaving} incorporates meta-cells that encode scenario information and introduces a meta-attention module to model inter-scenario correlations across two distinct tasks. Similarly, HiNet~\cite{zhou2023hinet} addresses Multi-scenario Multi-task modeling through hierarchical models that extract correlations both inter-domain and inter-task via a Scenario-aware Attentive Network~(SAN) and Customized Gate Control~(CGC) models. PEPNet~\cite{chang2023pepnet}, developed by Kuaishou, comprises two components: the Embedding Personalized Network~(EPNet), which personalizes embeddings to integrate features across domains, and the Parameter Personalized Network~(PPNet), designed to dynamically balance targets across multiple tasks. Additionally, models like 3MN~\cite{zhang20233mn} propose a scenario and task adaptive network for joint modeling, while AESM\textsuperscript{2} introduces a unified hierarchical structure for automatic expert selection, facilitating joint modeling in the Multi-scenario Multi-task context.
\section{Multi-Modal Recommendation}

In recommendation systems, joint modeling plays a crucial role, with representation extraction forming the bedrock of the overall framework. With the proliferation of multimedia services, item modalities have become increasingly diverse, encompassing images and texts among others. Consequently, the exploration of various modalities for representation extraction has garnered considerable attention due to its potential to address sparsity issues. This endeavor falls under the purview of multi-modal recommendation systems (MRS). Given the diverse nature of multi-modal data, where each modality resides in distinct semantic spaces, existing MRS literature predominantly focuses on the challenge of unifying modal representations into a cohesive space, a concept termed "feature interaction", which can be formulated as:
\vspace{-1mm}
\begin{equation}
    \boldsymbol{E}^M=M\left(\boldsymbol{E}^{t x t}, \boldsymbol{E}^v, ..., \boldsymbol{E}^p\right)=M\left(\mathcal{E}_{txt}(x^{txt}), \mathcal{E}_v(x^v),..., \mathcal{E}_p(x^p)\right)
\end{equation}
\noindent where $\textbf{E}^M$ is the unified representation and $x^*$ denote raw modality inputs. $\mathcal{E}_*(\cdot)$ represents the corresponding modality encoder, such as ViT~\cite{dosovitskiy2020image}. Thus, the core lies in the unifying function $M(\cdot)$. As depicted in Figure~\ref{fig_interaction}, these interactions are categorized into three main types: \textbf{Bridge}, \textbf{Fusion}, and \textbf{Filtration}. These techniques facilitate feature interaction from multiple viewpoints, enabling their simultaneous application within a single MRS model. For clarity, existing works are classified by interaction types in Table~\ref{tab:mrs}.

\begin{table}[t]
\centering
\caption{Category for Multi-modal Recommendation}
\resizebox{1\columnwidth}{!}{
\begin{tabular}{c|c|l}
\toprule[1.5pt]
\textbf{Interaction} & \textbf{Category} & \textbf{Related Works} \\ 
\midrule
\midrule
\multirow{3}{*}{Bridge} & User-item Graph & \cite{tao2020mgat}, \cite{ni2022two}, \cite{wang2021dualgnn}, \cite{wei2019mmgcn}, \cite{yi2022multi} \\
 & Item-item Graph & \cite{zhang2022latent}, \cite{zhang2021mining}, \cite{mu2022learning}, \cite{liu2021pre}, \cite{chang2020bundle}, \cite{ma2022crosscbr}, \cite{zhang2022latent} \\
 & Knowledge Graph & \cite{wang2020enhanced}, \cite{wang2019multi}, \cite{cao2022cross}, \cite{wang2022multimodal}, \cite{chen2022hybrid}, \cite{sun2020multi}, \cite{liu2022multi} \\ 
 \midrule
\multirow{5}{*}{Fusion} & Coarse-grained Attention & \cite{liu2022contrastive}, \cite{pan2022multimodal}, \cite{liu2019user}, \cite{chen2021cmbf} \\
 & \multirow{2}{*}{Fine-grained Attention} & \cite{chen2019personalized}, \cite{kim2022mario}, \cite{xiao2020deep}, \cite{lian2021multi}, \cite{tao2020mgat}, \cite{ni2022two}, \cite{liu2021pre}, \cite{chen2022hybrid}, \cite{han2022modality}, \cite{liu2022disentangled}, \cite{kim2022mario}, \\ & & \cite{liu2022implicit}, \cite{lei2021understanding}, \cite{liu2022multi}, \cite{chen2019pog}, \cite{hou2019explainable},\cite{lin2019explainable}, \cite{zhu2022combo}, \cite{wu2021mm}, \cite{li2022miner} \\
 & Combined Attention & \cite{liu2019nrpa}, \cite{liu2021noninvasive}, \cite{han2022vlsnr} \\
 & Other Fusion Methods & \cite{wang2021multimodal}, \cite{chen2021curriculum}, \cite{lv2019interest}, \cite{lv2019interest}, \cite{zhang2020multi}, \cite{xu2020recommendation} \\ 
 \midrule
Filtration & Filtration & \cite{sun2020multi}, \cite{zhou2022tale}, \cite{yinwei2021grcn}, \cite{liu2022megcf}, \cite{yi2021multi} \\ 
\bottomrule[1.5pt]
\end{tabular}
}
\label{tab:mrs}
\vspace{-5mm}
\end{table}

\subsection{Bridge}
Bridge focuses on capturing the inter-relationships between users and items while taking into account multi-modal information. This approach adopts the message-passing mechanism of GNNs to enhance user representation by facilitating information exchange between users and items, thereby capturing user preferences across different modalities. As illustrated in Figure~\ref{fig_interaction}, several works infer user $1$ preferences by aggregating interacted items for each modality. Furthermore, the modality representation of movie $1$ can be inferred from the latent item-item graph. In this subsection, we will delineate the methods for constructing bridges in MRS.

\subsubsection{\textbf{User-item Graph}}
By facilitating the exchange of information between users and items, it is feasible to capture users' preferences across various modalities. For example, MMGCN~\cite{wei2019mmgcn} derives a bipartite user-item graph based on the different modalities. Expanding upon MMGCN, GRCN~\cite{yinwei2021grcn} further improves the recommendation by dynamically adjusting the graph's structure while detecting the noisy interactions during the training. Despite their efficacy, these methods overlook variations in user preferences across different modalities. To tackle this issue, DualGNN~\cite{wang2021dualgnn} utilizes bipartite and co-occurrence graphs to extract the preferences from the user's correlations. Additionally, MMGCL~\cite{yi2022multi} introduces a novel multi-modal graph contrastive learning to address this challenge. MGAT~\cite{tao2020mgat} proposes a novel attention mechanism, facilitating the adaptive extraction of user preferences across various modalities.

\subsubsection{\textbf{Item-item Graph}}
Effectively leveraging item-item structures can also contribute to improving the learning of item representations, thereby enhancing model performance. For instance, approaches like LATTICE~\cite{zhang2021mining} and MICRO~\cite{zhang2022latent} establish item-item graphs for each modality by leveraging the user-item bipartite graph and amalgamating them to extract latent item graphs. Nonetheless, these methodologies overlook the nuances in preferences among specific user segments. As a remedy, HCGCN~\cite{mu2022learning} proposes a clustering graph convolutional network, which initially groups user-item and item-item graphs. Furthermore, drawing inspiration from the achievements of pre-training models, PMGT~\cite{liu2021pre} presents a pre-trained graph transformer, furnishing side information in a multi-modal type and a unified perspective on project relationships. In the context of bundle recommendation, models like BGCN~\cite{chang2020bundle} and Cross-CBR~\cite{ma2022crosscbr} amalgamate user-bundle, item-bundle, and user-item relationships into a heterogeneous graph.

\subsubsection{\textbf{Knowledge Graph}}
Many researchers have endeavored to integrate KGs with MRS by incorporating each modality of items into the KG as entities. As a pioneering work, MKGAT~\cite{sun2020multi} designs a multi-modal graph-based attention technique to extract information from multi-modal KGs, focusing on entity information extraction and relationship reasoning. Besides, MMKGV~\cite{liu2022multi} utilizes a graph attention network for disseminating and aggregating information on a multi-modal KG.

\begin{figure}[t]
    \centering
    \includegraphics[width=0.32\textwidth]{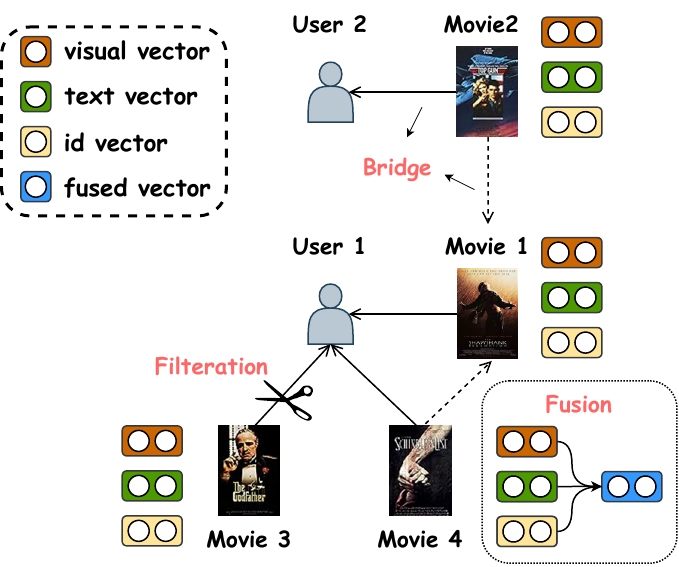}
    \vspace{-2.1mm}
    \caption{Three types of feature interaction for MRS.}
    \label{fig_interaction}
    \vspace{-5mm}
\end{figure}

\subsection{Fusion}
In contrast to bridge, fusion primarily revolves around exploring the intra-relationships among various modalities of items. Its main aim lies in merging preferences across different modalities. Among fusion techniques, the attention mechanism emerges as particularly prevalent, offering flexibility in integrating multi-modal information with varying emphasis and weights. In this subsection, as depicted in Figure~\ref{fig_interaction}, we begin by categorizing attention mechanisms based on their granularity of fusion before introducing alternative fusion approaches within MRS.

\subsubsection{\textbf{Coarse-grained Attention}}
Certain models utilize attention mechanisms to integrate information from multiple modalities at a coarse-grained level. For example, in UVCAN~\cite{liu2019user}, multi-modal information is segregated into user-side and item-side components. Information from the user side is then used to compute fusion weights for the item side through self-attention. MCPTR~\cite{liu2022contrastive} builds upon UVCAN by proposing parallel merging of item and user information. CMBF~\cite{chen2021cmbf} introduces a cross-attention mechanism to collectively capture semantic information from texts and images, followed by concatenation. Moreover, certain models exhibit variations in modal proportions, such as MML~\cite{pan2022multimodal} and MCPTR~\cite{liu2022contrastive}.

\subsubsection{\textbf{Fine-grained Attention}}
Multi-modal data encompasses both broad and nuanced features, such as audio tone variations and clothing patterns. Therefore, fine-grained fusion methods selectively integrate detailed feature information across various modalities. This fine-grained fusion is particularly crucial in fashion recommendation contexts. For instance, POG~\cite{chen2019pog} and NOR~\cite{lin2019explainable} employ multi-layer attention to extract deep features from fashion images' ensemble compositions, enabling continuous integration of intricate details. EFRM~\cite{hou2019explainable} enhances interpretability by incorporating a Semantic Extraction Network (SEN) to extract local features, which are then combined with fine-grained attention preferences. VECF~\cite{chen2019personalized} utilizes image segmentation techniques to merge image features from each patch with other modalities. Models like MINER~\cite{li2022miner}, DMIN~\cite{xiao2020deep}, and SUM~\cite{lian2021multi} create interest representations across various user aspects with the help of multi-modal information. Additionally, certain models~\cite{chen2022hybrid,tao2020mgat,kim2022mario} devise unique internal structures to enhance fine-grained fusion.

\subsubsection{\textbf{Combined Attention}}
Expanding on fine-grained fusion, specific models introduce combined fusion structures with the goal of preserving both fine-grained feature integration and global information aggregation. NOVA~\cite{liu2021noninvasive}, for example, proposes a non-invasive attention mechanism featuring two branches. It isolates embedding in a separate branch to retain interactive information during the fusion process. NRPA~\cite{liu2019nrpa} employs personalized word-level attention to identify crucial words in comments for each user/item, subsequently passing the comment information through alternating layers of fine and coarse-grained fusion. Similarly, VLSNR~\cite{han2022vlsnr}, models users' short and long-term interests, achieving both fusions by combining multi-head attention and GRU networks.

\subsubsection{\textbf{Other Fusion Methods}}
In addition to employing attention weights to fuse multi-modal information, some works utilize simpler techniques such as concatenation operations~\cite{zhang2020multi} and gating mechanisms~\cite{liu2021noninvasive}. However, these methods are rarely used independently and are often combined with graph and attention mechanisms, as previously discussed. Moreover, certain models fuse multi-modal features through both linear and nonlinear layers. Lv et al.~\cite{lv2019interest} incorporates a linear layer to merge textual and visual features. In MMT-Net~\cite{krishnan2020transfer}, three context invariants of restaurant data are labeled artificially and then fused through an MLP network.

\subsection{Filtration}
Multi-modal data differs from user interaction data as it often includes irrelevant information unrelated to user preferences. For example, interaction such as the one between movie $3$ and user $1$ shown in Figure~\ref{fig_interaction} may introduce noise and should thus be filtered out. Removing such noisy data in multi-modal recommendation tasks typically results in improved recommendation performance. Some works adopt image processing methods for denoising. For instance, VECF~\cite{chen2019personalized} and UVCAN~\cite{liu2022implicit} employ image segmentation to eliminate noise from images, thereby enhancing the capture of users' personalized interests. Similarly, MM-Rec~\cite{wu2021mm} employs target detection to identify significant image regions. Additionally, various structures based on GNNs are employed for denoising. FREEDOM~\cite{zhou2022tale} proposes a degree-sensitive edge pruning method to clean the user-item graph. GRCN~\cite{yinwei2021grcn} detects whether users interact with noisy items accidentally, while PMGCRN~\cite{jia2022preference} and MEGCF~\cite{liu2022megcf} address mismatched interactions. Moreover, MAGAE~\cite{yi2021multi} is devised to handle uncertainty issues for MRS.

\section{Multi-Behavior Recommendation}
Existing recommendation models usually focus solely on one single behavior of users as input data, such as clicking or purchasing items. However, in real-world recommendation scenarios, users tend to engage in various types of behavioral interactions with products.
For example, in video recommendation, users exhibit different behaviors towards a single video, such as clicks, likes, and retweets.
It is believed that additional behavioral information (also known as auxiliary behaviors) apart from the target behavior contains rich semantic information which can better assist in modeling user interests.
Consequently, multi-behavior recommendation (MBR) is applied in various commercial domains for more personalized and relevant recommendations based on multiple aspects of user behaviors.
To be specific, the key of MBR is to learn the representation of user and item under different behavior semantics
\begin{equation}
\begin{aligned}
&\boldsymbol{E}^{B}=G\left(\boldsymbol{H}_1, \boldsymbol{H}_2, \ldots, \boldsymbol{H}_N\right).
\end{aligned}
\end{equation}
Besides, there are several challenges MBR aims to address. The first is data sparsity \cite{li2022askme,chen2021graph,wei2022contrastive,wu2022multi,gu2022self,xu2023multi} since there are usually sparse supervision signal under the target behavior. The second is complex inter-behavior and intra-behavior relations \cite{xia2022multi2,chen2021curriculum}. The third is noise \cite{chen2021curriculum,zhang2023denoising} under auxiliary behaviors.
In the following, as shown in Table \ref{cmbr} the existing works are categorized into three groups: graph-based, transformer-based, and others.

\begin{table}[t]
\centering
\caption{Category for Multi-behavior Recommendation}
\label{cmbr}
\resizebox{1\columnwidth}{!}{
\begin{tabular}{c|l}
\toprule[1.5pt]
\textbf{Category} & \textbf{Related Works} \\ 
\midrule
\midrule
\multirow{2}{*}{Graph-based} & \cite{jin2020multi}, \cite{zhang2020multiplex}, \cite{wang2020beyond}, \cite{wang2021incorporating}, \cite{xiao2021dmbgn}, \cite{yang2021hyper}, \cite{chen2021graph}, \cite{xia2021multi}, \cite{xia2021graph}, \cite{wei2022contrastive}, \cite{wu2022multi} \\ 
 & \cite{gu2022self}, \cite{yan2022cascading}, \cite{meng2023coarse}, \cite{xia2022multi2}, \cite{weimulti}, \cite{zhang2023denoising}, \cite{cheng2023multi}, \cite{xu2023multi}, \cite{yan2023mb}, \cite{li2023dual}, \cite{rang2023heterogeneous}, \cite{yan2023cascading}\\
\midrule
Transformer-based & \cite{guo2019buying}, \cite{xia2020multiplex}, \cite{gu2020deep}, \cite{li2022askme}, \cite{ano2022multi}, \cite{wu2021feedrec} \\
\midrule
Others & \cite{gao2019neural}, \cite{chen2020efficient}, \cite{zhao2020difference}, \cite{chen2021curriculum}, \cite{wang2022causal}, \cite{liu2023modeling}, \cite{gan2023multi}, \cite{li2023dual}, \cite{cai2024robust},\cite{xia2021knowledge}, \cite{xia2022multi}, \cite{yang2022multi} \\ 
\bottomrule[1.5pt]
\end{tabular}
}
\label{tab:interaction}
\vspace{-5mm}
\end{table}

\subsection{Graph-based Methods}
Most existing works on multi-behavior recommendation adopt graph neural network in their framework design. 
The reason is that the performance of graph-based methods is generally better than that of non-graph models. A possible explanation is that the information propagation and aggregation based on the graph structure can better mine the complex semantic relationships of heterogeneous behaviors of users.

Generally, current models transform users' heterogeneous historical interactions into multi-behavior interaction graphs, where each user $u$ and item $i$ is a node, and there is a behavior-specific edge between them if $u$ interacts with $i$ under some behavior. For example, MBGCN \cite{jin2020multi} conducts behavior-aware user-item propagation and item-relevance aware item-item propagation in the user-item graph.
Meanwhile, another significant technique leveraged by many graph-based methods is contrastive learning, which is a kind of self-supervised learning paradigm and is able to learn more distinguishable representations. Specifically, different behaviors can be treated as different contrastive views \cite{wei2022contrastive,wu2022multi,weimulti}. Nevertheless, it may also suffer from low efficiency because it requires introducing random factors like graph augmentation operation. 

\subsection{Transformer-based Methods}

To begin with, according to \cite{yang2022multi}, BERT4Rec \cite{sun2019bert4rec} can be enhanced to tackle MBR by injecting behavior representations into input for self-attention. Afterward, following the transformer architecture for sequential recommendation, for example, DMT \cite{gu2020deep} employs multiple deep interest Transformers to model different behavioral sequences, representing users' real-time interests with multiple low-dimensional vectors and making multi-task predictions. Compared to traditional Transformer structures, MB-STR \cite{ano2022multi} designs multiple behavior Transformer layers to capture the heterogeneous dependencies among multiple behaviors and the unique semantics of behaviors simultaneously. However, they usually achieve inferior performance than graph-based methods.

\subsection{Other Methods}

In the early stage of MBR, some works simply regard different behaviors as different tasks to be predicted and adopt a multi-task learning framework \cite{gao2019neural,chen2020efficient}. This idea is closely related to the cascaded task relation introduced in Section \ref{casc} which supposes the sequential relationship among behaviors as a prior. For example, EHCF \cite{chen2020efficient} leverages a transfer-based prediction and proposes an efficient optimization method without the need of sampling. However, it is difficult for these methods to comprehensively depict and understand the complex cross-type behavior dependencies in real-world recommendation scenarios.

Thanks to the strong capability of representing and capturing complex relations between nodes, some methods adopt graph transformer to model multi-behavior dependencies \cite{xia2021knowledge,xia2022multi,yang2022multi}. For example, KHGT \cite{xia2021knowledge} proposes a hierarchical graph transformer network with the graph-structured transformer module and attentive fusion network to capture high-order relations in the knowledge-aware multi-behavior graph. However, they also face similar challenges of graph- and transformer-based methods.

\section{Future Directions}


\textbf{Multi-task Recommendation}.   As discussed in Section \ref{nt}, previous works try to tackle negative transfer from either gradient or separating shared and specific parameters. However, how to extract the complex inter-task correlation needs further research, e.g., from the causal relation \cite{chen2022cfs}. Meanwhile, what, where, and when to transfer to alleviate negative transfer is still under-explored.

\noindent \textbf{Multi-scenario Recommendation}. With scenarios numbering in the thousands, there is a pressing need for scenario-unified models optimized for real-time performance. This necessitates further exploration into quantification~\cite{ko2021mascot} and compression~\cite{yin2021tt}. Furthermore, with scenario interrelations becoming complex, combining LLMs with MSR models to extract scenario information and bridge semantic gaps poses an urgent challenge.

\noindent \textbf{Multi-modal Recommendation}. Despite the availability of methods proposed for different interaction types within a model~\cite{lei2021understanding}, the absence of an up-to-date universal solution that integrates these techniques remains a notable gap. 
Besides, the current dataset for MRS is rather constrained, lacking comprehensive coverage of various modalities, especially for joint modeling.

\noindent \textbf{Multi-behavior Recommendation} Existing works, especially Graph-based methods necessitate a full incremental update pattern, where aggregation operations are performed on all nodes at each layer of GNN, resulting in a significant computational cost. How to tackle their scalability and deployment in large-scale industrial recommender systems requires further exploration.


\noindent \textbf{Other Joint Modeling Perspective}. 
First, instead of representing user interests with a single vector, multi-interest recommendation adopts multiple representations to accurately capture the user's dynamic and diverse preferences \cite{li2019multi,cen2020controllable,pi2019practice,tian2022multi,li2022improving}.
Second, based on multi-objective optimization, multi-objective recommendation \cite{jannach2022multi,zheng2022survey} focuses on the trade-off and balance among objectives, e.g., diversity \cite{liu2021diversity} and fairness \cite{xiao2017fairness} from the optimization perspective.

\noindent \textbf{Novel Joint Modeling Technology}. 
Traditional DRSs are designed to meet specific requirements. However, the use of Large Language Models (LLMs) introduces a new opportunity for improvement due to their strong understanding of language and context. Currently, LLMs are mainly used to either add context~\cite{acharya2023llm,wei2024llmrec} or make recommendations simply based on fine-tuning~\cite{bao2023tallrec,geng2022recommendation}. Nevertheless, the real potential lies in fully integrating LLMs with multiple joint modeling dimensions to improve all aspects of recommender systems. Therefore, exploring how to effectively merge LLMs with joint modeling methods offers a promising direction for future research, aiming for more versatile and effective systems.
\section{Conclusion}
This survey highlights the potential of joint modeling in DRSs, addressing the limitations of traditional approaches by integrating multi-task, multi-scenario, multi-modal and multi-behavior strategies. This comprehensive approach promises enhanced personalization, efficiency, and user satisfaction. The exploration of joint modeling's various dimensions and its challenges and opportunities underscores the complexity of developing more sophisticated recommendation platforms. Looking forward, the advancement of joint modeling techniques is crucial for the evolution of recommender systems with research directions focusing on algorithm refinement and scalability. This survey aims to catalyze future innovation, guiding the development of next-generation DRSs that are more adaptive, intelligent, and user-centric.

\bibliographystyle{ACM-Reference-Format}
\bibliography{newbibfile}

\end{document}